\documentclass[fleqn]{article}
\usepackage{gc,amssymb}

\def\GR{general relativity}
\def\Sch{Schwarzschild}
\def\dS{de Sitter}

\def\sph{spherically symmetric}
\def\ssph{static, spherically symmetric}

\def\bh{black hole}
\def\bhs{black holes}
\def\wh{wormhole}
\def\whs{wormholes}
\def\bw{brane world}

\def\asflat{asymptotically flat}

\def\thd{\fract 13}
\def\qua{\fract 14}
\def\mn{_{\mu\nu}}
\def\MN{^{\mu\nu}}
\def\mN{_\mu^\nu}

\def\cF{{\cal F}}
\def\dens{${\rm g\, cm}^{-3}$}

\heads{K.A. Bronnikov, S.B. Fadeev and A.V. Michtchenko}
{Scalar fields in a minimally coupled brane world:
	no-hair and other no-go theorems}

\begin{document}
\thispagestyle{empty}
\prepno{gr-qc/0301106}{}
\vspace*{-4mm}

\Title {Scalar field in a minimally coupled brane world: \yy
	no-hair and other no-go theorems}

\Aunames{K.A. Bronnikov\auth{a,b,1},
         S.B. Fadeev\auth{a} and A.V. Michtchenko\auth{c,2}}

\Addresses{
\addr a {VNIIMS, 3-1 M. Ulyanovoy St., Moscow 117313, Russia}
\addr b {Institute of Gravitation and Cosmology, PFUR,
        6 Miklukho-Maklaya St., Moscow 117198, Russia}
\addr c {SEPI--ESIME, IPN, Zacatenco, M\'exico, D.F., CP07738, Mexico}
}

\Abstract
{In the brane-world framework, we consider static, spherically symmetric
configurations of a scalar field with the Lagrangian $(\d\phi)^2/2 -
V(\phi)$, confined on the brane. We use the 4D Einstein equations on the
brane obtained by Shiromizu et al., containing the conventional stress
tensor $T\mN$, the tensor $\Pi\mN$ which is quadratic in $T\mN$, and $E\mN$
describing interaction with the bulk. For models under study, the tensor
$\Pi\mN$ has zero divergence, allowing one to consider $E\mN =0$. Such a
brane, whose 4D gravity is decoupled from the bulk geometry, may be called
minimally coupled. Assuming $E\mN =0$, we try to extend to brane worlds
some theorems valid for scalar fields in general relativity (GR). Thus, the
list of possible global causal structures in all models under consideration
is shown to be the same as is known for vacuum with a cosmological constant
in GR:  Minkowski, Schwarzschild, (anti-)de Sitter and Schwarzschild--%
(anti-)de Sitter. A no-hair theorem, saying that, given a potential $V\geq
0$, asymptotically flat black holes cannot have nontrivial external scalar
fields, is proved under certain restrictions. Some objects, forbidden in GR,
are allowed on the brane, e.g, traversable wormholes supported by a scalar
field, but only at the expense of enormous matter densities in the strong
field region.}

\email 1 {kb@rgs.mccme.ru}
\email 2 {mial@maya.esimez.ipn.mx}

\section{Introduction}

    The widely discussed brane-world models suppose that the observable
    Universe is a kind of domain wall, or brane, in a multidimensional
    space, with large or even infinite extra dimensions. The standard-model
    fields are supposed to be confined on the brane while gravity propagates
    in the surrounding bulk. The history of such models is traced back at
    least to the early 80s \cite{early}, cetrain progress then followed (see,
    e.g., \cite{later}), but the recent outburst of interest owes to the
    progress in string theory and M-theory, in particular, to the
    Horava-Witten 11-dimensional model \cite{ho-wit}. The brane-world
    approach has suggested a natural mechanism of solving the hierarchy
    problem of particle physics \cite{dvali,RS1,RS2}, preserving a
    correct behaviour of weak gravity on the brane. Many related important
    problems of cosmology and particle physics have also been addressed, see
    the recent reviews \cite{brane-rev}.

    Most of modern studies consider a simplified version of this picture,
    the so-called RS2 class of models \cite{RS2} gravity acts in a
    5-dimensional bulk while our 4-dimensional world is a time-evolving
    3-brane which supports all standard-model fields. Gravity on the brane
    is described by modified 4-dimensional Einstein equations \cite{maeda99}
    [\eqs (\ref{EE4}) in \sect 2] which contain additional terms leading to
    many new predictions as compared to \GR\ (GR). One such term, the
    traceless tensor $E\mN$, originating from the 5D Weyl tensor and
    describing tidal effects on the brane from the bulk geometry, may
    contribute to ``dark energy'' in cosmology \cite{brane-rev} and leads to
    numerous new \bh\ \cite{bra-bh,bbh1} and \wh\ \cite{br-kim} solutions on
    the brane.  Another additional term, $\Pi\mN$, which is quadratic in the
    stress-energy tensor (SET) of matter confined on the brane, contributes
    to the cosmological evolution \cite{brane-rev} and can lead to new
    strong field effects in local gravitational fields. Like other models,
    brane-world cosmologies often involve scalar fields with various
    potentials \cite{brane-rev,bra-scalar,shta}, which strongly affect the
    cosmological dynamics and are probably necessary for describing
    inflation and dark energy. Less is known about the properties of scalar
    fields in the physics of local objects on the brane.

    In this paper, we consider the modified Einstein equations for \ssph\
    configurations of scalar fields with arbitrary potentials $V(\phi)$ in
    the absence of other kinds of matter (scalar-vacuum configurations, for
    short) and try to extend to brane worlds some theorems valid for scalar
    fields in GR. For minimally coupled scalar fields described by the
    Lagrangian $L_s = \half (\d\phi)^2 - V(\phi)$, among such theorems are:

\begin{description}    \itemsep 1pt
\item[(i)] The no-hair theorems \cite{bek,ad-pear}: in case $V\geq 0$,
    as well as for some partly negative potentials, an \asflat\ \bh\ cannot
    have a nontrivial external scalar field.

\item[(ii)] The ``generalized Rosen theorem'' saying that particlelike
    scalar-vacuum solutions (\asflat\ solutions with a regular centre) are
    absent as long as $V(\phi) \geq 0$ (see \cite{vac5} and references
    therein).

\item[(iii)] The nonexistence theorem for regular solutions without a
    centre (e.g., wormholes) \cite{vac1}.

\item[(iv)] The causal structure theorem \cite{vac1}, asserting that
    the list of possible types of global causal structures (and the
    corresponding Carter-Penrose diagrams) for scalar field configurations
    with any potentials $V(\varphi)$ and any spatial asymptotics is the same
    as is known for vacuum with a cosmological constant in GR.
\end{description}

    Extensions of these theorems to scalar-tensor and Kaluza-Klein type
    multidimensional theories have been considered before
    \cite{vac2,vac3,vac4,vac5}. For scalar vacuum in a \bw, the situation is
    complicated by the non-closed nature of the 4D equations (\ref{EE4}):
    the ``tidal'' tensor $E\mN$ bears information on the bulk gravity;
    moreover, in the general case there is stress-energy exchange between
    $E\mN$ and matter on the brane represented by the quadratic tensor
    $\Pi\mN$. Fortunately, for a scalar field in the \ssph\ case $\Pi\mN$ is
    conservative, making it possible to put $E\mN \equiv 0$.  This
    assumption closes the set of 4D field equations. Such a brane may be
    called {\it minimally coupled\/}: 4-dimensional gravity on the brane is
    then decoupled from the bulk geometry. For $T\mN=0$, the 4D gravity
    equations reproduce the vacuum Einstein equation, and, for $T\mN\ne 0$,
    the only possible brane-world effect is connected with the tensor
    $\Pi\mN$, quadratic in $T\mN$ and significant in strong fields.

    A full description of the model evidently requires knowledge of the
    whole 5D geometry. Any specific solution on the brane must be
    ``evolved'' into the bulk, which is a nontrivial problem even for very
    simple brane metrics \cite{bra-bh,casad02,wiseman}. We will not discuss
    the possible bulk properties of models in question and only note that
    the existence of the corresponding solutions to the higher-dimensional
    equations of gravity (in our case, the 5D vacuum Einstein equations with
    a cosmological term) is guaranteed by the Campbell-Magaard type
    embedding theorems \cite{campmaag}. A recent discussion of these
    theorems, applied, in particular, to \bw\ scenarios, and more references
    can be found in Ref.\,\cite{wess03}.

    In what follows, considering scalar fields on a minimally coupled brane,
    we shall see that the above theorems do not always hold even in this
    idealized case. The causal structure theorem proves to be valid without
    change. Bekenstein's \cite{bek} no-hair theorem for scalar fields with
    potentials $V = V(\phi^2)$ such that $dV/d(\phi^2) \geq 0$ also remains
    valid. On the contrary, solitonic (particlelike) solutions for $V\geq 0$
    as well as wormholes cannot be excluded. Wormhole throats are admissible
    in a strong field region, and this circumstance does not allow us to
    prove the no-hair theorem for scalar fields with arbitrary nonnegative
    potentials in its full generality: one cannot exclude a \bh\ with scalar
    hair whose event horizon is located beyond a throat. However, an
    estimate based on an experimental restriction on the bulk length scale
    shows that both \wh\ throats and particlelike solutions require energy
    densities and pressures far beyond the nuclear density.

\section{Field equations and conservation laws}

    The gravitational field on the brane is described by
    the modified Einstein equations derived by Shiromizu, Maeda and Sasaki
    \cite{maeda99} from 5-dimensional gravity with the aid of the Gauss and
    Codazzi equations\footnote
{The sign conventions are: the metric signature $(+{}-{}-{}-)$; the curvature
 tensor $R^{\sigma}_{\ \mu\rho\nu} = \d_\nu\Gamma^{\sigma}_{\mu\rho}-\ldots$,
 so that, e.g., the Ricci scalar $R > 0$ for de Sitter space-time, and the
 stress-energy tensor (SET) such that $T^t_t$ is the energy density.}:
\bear
    G\mN = - \Lambda_4\delta\mN -\kappa_4^2 T\mN
	         - \kappa_5^4 \Pi\mN - E\mN,                     \label{EE4}
\ear
    where $G\mN = R\mN - \half \delta\mN R$ is the 4D Einstein tensor,
    $\Lambda_4$ is the 4D cosmological constant expressed in terms of
    the 5D cosmological constant $\Lambda_5$ and the brane tension $\lambda$:
\beq
     \Lambda_4 = \Half
     \biggl(\Lambda_5 + \frac{1}{6} \kappa_5^4\lambda^2\biggr);  \label{La4}
\cm
    \kappa_4^2 = 8\pi G_{\rm N} = \kappa_5^4 \lambda/(6\pi),
\eeq
    where $\kappa_4$ is the 4D gravitational constant, $G_{\rm N}$ is
    Newton's constant of gravity; $T\mN$ is the SET of matter confined on
    the brane; $\Pi\mN$ is a tensor quadratic in $T\mN$, obtained from
    matching the 5D metric across the brane:
\beq                                                          \label{Pi_}
    \Pi\mN = \fract{1}{4} T_\mu^\alpha T_\alpha^\nu - \half T T\mN
	     - \fract{1}{8} \delta\mN
		\left( T_{\alpha\beta} T^{\alpha\beta} -\thd T^2\right)
\eeq
    where $T = T^\alpha_\alpha$; lastly, $E\mN$ is the ``electric'' part of
    the 5D Weyl tensor projected onto the brane: in proper 5D coordinates,
    $E\mn = \delta_\mu^A \delta_\nu^C {}^{(5)} C_{ABCD} n^B n^D$ where
    $A, B, \ldots$ are 5D indices and $n^A$ is the unit normal to the brane.
    By construction, $E\mN$ is traceless, $E_\mu^\mu = 0$ \cite{maeda99}.

    The contracted Bianchi identities $\nabla_\nu G\mN =0$ lead to the
    corresponding relation for the right-hand side of (\ref{EE4}).
    Assuming the conservation law $\nabla_\nu T\mN =0$ for matter,
    we obtain the condition
\beq
    \nabla_\nu (\Pi\mN + E\mN) =0,                           \label{cons+}
\eeq
    but $\Pi\mN$ and $E\mN$, taken separately, are not necessarily
    divergence-free: there is, in general, stress-energy exchange between
    the brane matter and the bulk gravity.

    Consider now a minimally coupled scalar field confined on the brane, with
    the 4D Lagrangian
\beq
      L_s = \half (\d\phi)^2 - V(\phi),                       \label{L_s}
\eeq
    where $(\d\phi)^2 = g\MN \phi_{,\mu}\phi_{,\nu}$. The field equation and
    the tensor $T\mN$ have the conventional form
\bearr
    \nabla^\nu \nabla_\nu\phi + dV/d\phi =0,                  \label{Eq_s}
\\ \lal
    T\mN = \phi_{,\mu} \phi^{,\nu} -\half \delta\mN (\d\phi)^2
		+ \delta\mN V.
\ear
    The tensor $T\mN$ is conservative while the corresponding $\Pi\mN$ is,
    in general, not:
\bear                                                           \label{Pi}
    \Pi\mN \eql \phi_{,\mu}\phi^{,\nu}
	     \left[ \fract{5}{12}(\d\phi)^2 - \fract{7}{6} V \right]
       + \delta\mN  \left[ -\fract{1}{16}(\d\phi)^4
		-\fract{5}{12}V(\d\phi)^2 +\fract{17}{12} V^2 \right];
\yy
   \nabla_\nu \Pi\mN \eql \fract{1}{3} \phi^{,\nu} \phi^{,\alpha} \label{dPi}
    	    (\phi_{,\mu} \phi_{;\alpha\mu} - \phi_{,\nu}\phi_{;\alpha\mu})
     + \fract{1}{4}V \phi^{,\alpha}\phi_{;\alpha\mu}
			+ \half V \phi_{,\mu} \nabla^\nu \nabla_\nu\phi.
\ear

    For \ssph\ configurations, the brane metric can be written in the form
\beq                                                             \label{ds}
     ds^2 = A(\rho) dt^2 - \frac{d\rho^2}{A(\rho)} - r^2(\rho)d\Omega^2,
\eeq
    where $d\Omega^2 =d\theta^2 + \sin^2 \theta\,d\varphi^2$ is the linear
    element on a unit sphere and $\rho$ is the radial coordinate under the
    convenient ``quasiglobal'' gauge $g_{00}g_{11} = -1$. With this metric,
    the scalar field $\phi(\rho)$ has the SET
\beq
     T\mN = \half \diag (f, -f, f, f) + \delta\mN\, V(\phi), \cm
	  f \eqdef A(\rho) \phi'^2,                           \label{T-sph}
\eeq
    where the prime denotes $d/d\rho$. The corresponding $\Pi\mN$ has the
    nonzero components
\beq                                                        \label{Pi-sph}
     \Pi_0^0 = \Pi_2^2 = \Pi_3^3 = \frac{1}{48} (4V^2 + 4fV -3f^2);
     \cm \Pi_1^1 = \frac{1}{48} (4V^2 - 4fV + f^2),
\eeq
    and a direct inspection leads to the important conclusion that {\it
    the tensor $\Pi\mN$ is conservative in the \ssph\ case\/}:
    $\nabla_\nu \Pi\mN =0$.

    This means, in particular, that the assumption $E\mN =0$ does not
    contradict the field equations for scalar fields on the brane. Under
    this assumption, gravity on the brane decouples from the bulk gravity,
    and \eqs (\ref{EE4}), (\ref{Eq_s}) comprise a closed set of
    equations. They can be written as follows:
\bear
       (Ar^2 \phi')' \eql r^2 dV/d\phi;                      \label{phi}
\\
       \frac{1}{2r^2} (A' r^2)' \eql
	  	-\Lambda_4 - \kappa_4^2 V
	       + \frac{\kappa_5^4}{48} (4V^2 - A^2 \phi'^4); \label{00}
\\                                                           \label{01}
       2 \frac{r''}{r} \eql - \kappa_4^2 {\phi'}^2
       \biggl[ 1 + \frac{\kappa_5^4}{12 \kappa_4^2}(2V - A\phi'^2)\biggr];
\\
       A (r^2)'' - r^2 A'' \eql 2 .                          \label{02}
\\                                                           \label{int}
      \frac{1}{r^2} (-1 + A'rr' + Ar'^2) \eql -\Lambda_4
      		       + \Half \kappa_4^2 (A\phi'^2 - 2V)
		       - \frac{\kappa_5^4}{48} (A\phi'^2 - 2V)^2.
\ear
    Only three of the five equations are independent, in particular,
    (\ref{int}) is a first integral of (\ref{00})--(\ref{02}). Given a
    potential $V(\varphi)$, this is a determined set of equations for the
    unknowns $r,\ A,\ \varphi$.

    The scalar-vacuum equations of GR are restored in case $\kappa_5=0$.

\section{No-go theorems} 

    Let us try to extend some theorems known in GR to a minimally coupled
    \bw, using \eqs (\ref{phi})--(\ref{int}).

\subsection{Causal structure theorem}  

    The first theorem concerns the possible number and order of Killing
    horizons, coinciding with the number and order of zeros of $A(\rho)$ at
    finite $r$. A simple (first-order) or any odd-order horizon separates a
    static region, $A > 0$ (also called an R region), from a nonstatic
    region, $A < 0$, where (\ref{ds}) is a homogeneous cosmological metric
    of Kantowski-Sachs type (a T region). A horizon of even order separates
    regions with the same sign of $A(\rho)$. The gauge $g_{00} g_{11}=-1$
    makes it possible to jointly consider regions on both sides of a horizon
    since, as is directly verified, $\rho$ behaves in its neighbourhood like
    a manifestly regular Kruskal-like coordinate.

    The disposition of horizons unambiguously determines the global causal
    structure of space-time, up to identification of isometric surfaces,
    if any. The following theorem severely restricts such possible
    dispositions.

\Theorem{Theorem 1}
    {Consider solutions to \eqs (\ref{phi})--(\ref{int}).
     Let there be a static region $a < \rho < b \leq \infty$. Then:
\begin{description}\itemsep -2pt
\item [(i)]
     all horizons are simple;
\item [(ii)]
     no horizons exist at $\rho < a$ and at $\rho > b$.
\end{description} }
\vspace{-1ex}

     A proof of this theorem is the same as in \Ref{vac1} and rests on
     the property $T_0^0 = T_2^2$ of the SET (\ref{T-sph}), leading to
     the same property of the tensor $\Pi\mN$. The idea is that
     \eq (\ref{02}), expressing the equality $R_0^0 = R^2_2$, can be
     rewritten as
\beq
     r^4 B'' + 4 r^3 r'B' = -2                               \label{02'}
\eeq
     where $B(\rho) = A/r^2$. At points where $B'=0$, we have $B''< 0$,
     therefore $B(\rho)$ cannot have a regular minimum. So, having once
     become negative while moving to the left or to the right along the
     $\rho$ axis, $B(\rho)$ (and hence $A(\rho)$) cannot return to zero or
     positive values.

     By Theorem 1, there can be at most two simple horizons around a static
     region. A second-order horizon separating two nonstatic regions can
     appear, but this horizon is then unique, and the model has no static
     region.

     The possible dispositions of zeros of the function $A(\rho)$, and hence
     the list of possible global causal structures, are thus the same as in
     the case of \sph\ vacuum in GR with a cosmological constant. The types
     of Carter-Penrose diagrams that are possible with solutions to \eqs
     (\ref{phi})--(\ref{02}) are exhausted by the diagrams known for
     different particular cases of the \Sch--\dS\ solution, described by
     the metric (\ref{ds}) with $r \equiv \rho$ and $A = 1 - 2m/r -
     \Lambda_4 r^2/3$. The only structure types are Minkowski, \Sch, \dS,
     AdS, and \Sch-(A)dS, see Refs. \cite{SdS,vac2,vac3,vac4} for more
     detail. Asymptotically flat \bhs, if any, have the \Sch\ structure.

     Theorem 1 is independent of any assumptions on the shape of the
     potential and on the spatial asymptotics.

\subsection{Theorems which fail in a \bw}

     The absence of \wh\ throats (i.e., regular minima of $r$) for
     scalar-vacuum configurations in GR follows from \eq (\ref{01}) which,
     in case $\kappa_5 = 0$, leads to $r'' \leq 0$, so that a regular
     minimum of $r(\rho)$ is impossible. However, in case $\kappa_5\ne 0$ the
     r.h.s. of \eq (\ref{01}) can become positive in a strong-field region,
     therefore \wh\ throats and \whs\ as global scalar-vacuum solutions on
     the brane are not excluded.

     Consider now \eqs (\ref{phi})--(\ref{int}) in case $\Lambda_4 = 0$,
     which is a natural assumption when dealing with local objects, very
     small on the cosmological scale. It is then natural to study \asflat\
     configurations. A flat asymptotic $\rho\to \infty$ is characterized by
\beq
     r \approx \rho, \cm
     		A = 1-2G_{\rm N} m/\rho + o(\rho^{-1}),	     \label{flatas}
     	\cm
     \phi = o(\rho^{-3/2}), \cm    V(\phi) = o(\rho^{-3}),
\eeq
     where $m$ is the \Sch\ mass. According to \eq (\ref{phi}), at a flat
     asymptotic $dV/d\phi = o(\rho^{-5/2})$, so that both $V$ and $dV/d\phi$
     vanish at infinity. These are minimal requirements that follow directly
     from the field equations. If, in addition, we assume that $V(\phi)$ is
     at least $C^2$-smooth and $\phi$ can be expanded in powers of $1/\rho$,
     it follows that $dV/d\phi$ must vanish at large $\rho$ as $\rho^{-4}$ or
     more rapidly.

     The nonexistence theorem for particlelike solutions (i.e., \asflat\
     solutions with a regular centre) follows for scalar vacuum in GR from a
     universal identity \cite{vac5}, valid for any \ssph\ particlelike
     solution and obtained by comparing two expressions for the mass $m$.
     One of them is the standard integral $m = 4\pi G\int T_0^0 r^2 dr$, the
     other follows from Tolman's well-known formula, and their comparison
     leads to the identity
\beq                                                         \label{univ}
     \int_{\rho_c}^{\infty}
          \bigl [(r'-1) T_0^0 + T^1_1 + T^2_2 + T^3_3 \bigr ] r^2 d\rho =0,
\eeq
     where $\rho_c$ is the value of $\rho$ at the centre. Applied to a
     scalar field with the SET (\ref{T-sph}), \eq (\ref{univ}) takes the
     form
\beq                                                        \label{univ-s}
     \int_{\rho_c}^{\infty}
                    \bigl[ f + 2V (2 + r') \bigr] r^2 d\rho =0.
\eeq
     In GR we always have $r'(\rho) \geq 1$ due to $r'(\infty) = 1$ and $r''
     \leq 0$, hence (\ref{univ-s}) cannot be satisfied without assuming
     $V < 0$ at least in some range of $\rho$.

     For a particlelike solution on the brane the identity (\ref{univ})
     is also valid but with $T\mN$ replaced by $T\mN + \Pi\mN$, and instead
     of (\ref{univ-s}) we obtain
\beq                                                        \label{univ-b}
     \int_{\rho_c}^{\infty} \biggl\{\kappa_4^2
               			[ f + 2V (2 + r') \bigr]
     + \frac{1}{24}\kappa_5^4
	      \bigl [(f - 2V)^2 r' + 8V^2 - 2f^2]\biggr\} r^2 d\rho = 0.
\eeq
     As we saw, $r(\rho)$ now can have a minimum, so we cannot rule out
     $r'< 0$ and even $r' < -2$ in a certain region of space; moreover, the
     integrand in (\ref{univ-b}) contains the term $- 2f^2$. Therefore
     (\ref{univ-b}) can hold with nonnegative $V$.

\subsection{No-hair theorems}

     Let us continue studying \asflat\ solutions and try to extend the
     no-scalar-hair theorem to brane worlds.

     First of all we notice that \eq (\ref{phi}) is the same as in GR,
     therefore Bekenstein's argument \cite{bek} that rests solely on the
     scalar field equation in a \bh\ geometry and does not employ the
     equations of gravity, is applicable here without change and can only be
     slightly refined in our \sph\ case due to Theorem 1. Let us reproduce
     it in our notation.

\Theorem{Theorem 2}
     {Given a potential $V = V(\phi^2)$ such that $dV/d(\phi^2) \geq 0$,
     the only \asflat\ \bh\ solution to \eqs (\ref{phi})--(\ref{02})
     is characterized by $V\equiv 0$, $\phi = \const$ and the \Sch\ metric
     in the whole domain of outer communication.}

     \noi
     {\bf Proof.}
     Let $\rho=h$ be the event horizon, so that $A(h) =0$ and $r(h)$ is
     finite, and let $\rho\to \infty$ be a flat asymptotic, so that
     $A(\rho) > 0$ in the range $h < \rho < \infty$.  Multiplying \eq
     (\ref{phi}) by $\phi$, we can rewrite it in the form
\beq
     \bigl(r^2 A \phi'^2\bigr)' - r^2 A \phi'^2 = r^2 \phi\frac{dV}{d\phi},
\eeq
     which gives after integration from $h$ to infinity:
\beq                                                           \label{int1}
     \int_{h}^{\infty} \biggl(A\phi'^2  + \phi\, \frac{dV}{d\phi}\biggr)
     		r^2 d\rho = r^2 A\phi\phi' \Big |_h ^\infty.
\eeq
     At the flat asymptotic, according to (\ref{flatas}), $\phi\phi' =
     o(r^{-2})$, so that $r^2 A\phi\phi' \Big |^{\rho\to\infty} =0$.
     Furthermore, as $\rho \to h$, $A \sim \rho-h$ since, by
     Theorem 1, it is a simple horizon. A horizon is a regular surface, at
     which $f=A\phi'^2$ must be finite (otherwise the components of $T\mN$
     would blow up, leading, due to gravity equations, to a singular
     geometry). Therefore, even if $\phi'$ blows up as $\rho\to h+0$, it
     cannot grow faster than $(\rho-h)^{-1/2}$, so that $|\phi(h)| <
     \infty$.  As a result, we have $A\phi\phi'=0$ at $\rho=h$, the whole
     r.h.s. of \eq (\ref{int1}) is zero, making the integral in the l.h.s.
     vanish.

     By assumption, $\phi V_\phi = 2\phi^2 dV/d(\phi^2) \geq 0$, so the
     integrand in (\ref{int1}) is nonnegative and, to produce a zero
     integral, it must vanish identically for $h < \rho < \infty$: we must
     have $\phi=\const$ and (since $V \to 0$ at the asymptotic) $V\equiv 0$.
     $\DAL$

\medskip
     Note that Theorem 2 did not need the condition $V\geq 0$, but only
     $dV/d(\phi^2) \geq 0$. The theorem even holds for some partly negative
     potentials. A good example is $V = K(\phi^2 - \eta^2)^3$ with positive
     constants $K$ and $\eta$. Indeed, both $V$ and $dV/d\phi$ vanish at
     $\phi=\pm \eta$, making $V(\phi)$ compatible with a flat asymptotic;
     $dV/d(\phi^2) = 3 K(\phi^2-\eta^2)^2 \geq 0$, the condition of Theorem
     2 is satisfied; and $V < 0$ at $-\eta < \phi < \eta$.

     In \Ref{bek} Theorem 2 was proved in a more general setting, for static
     \asflat\ \bhs\ without the spherical symmetry assumption.

     Now, assuming a general nonnegative potential, let us prove the
     following theorem, using an argument somewhat similar to that of Adler
     and Pearson \cite{ad-pear}.

\Theorem{Theorem 3}
    {Given a potential $V(\phi)\geq 0$, the only \asflat\ \bh\ solution to
     \eqs (\ref{phi})--(\ref{02}), such that $r' > 0$ at and outside the
     event horizon, is characterized by $\phi = \const$ and the \Sch\ metric
     in the whole domain of outer communication.}

     \noi
     {\bf Proof.}
     Let again $\rho = h$ be the \bh\ event horizon and $h < \rho < \infty$
     the domain of outer communication. Consider the function
\beq
      \cF_1(\rho) = \frac{r^2}{r'}[ 2V - A \phi'^2 ],         \label{F1}
\eeq
     which is everywhere meaningful due to our assumption $r' >0$.
     Let us find $\cF'_1$, substituting $\phi''$ from (\ref{phi}), $r''$
     from (\ref{01}) and $A'$ from (\ref{int}). The result is
\bearr
      \cF_1'(\rho) = \cF_2(\rho) \eqdef                           \label{F2}
        r\biggl[4 V +  \frac{\phi'^2}{r'^2} + A\phi'^2
		    + \frac{\kappa_5^2}{48} \frac{r^2\phi'^2}{r'^2}
			(2V - A \phi'^2)^2 \biggr],
\ear
     or, after integration from $h$ to infinity,
\beq                                                            \label{iden}
     \cF_1(\infty) - \cF_1(h) = \int_{h}^{\infty} \cF_2 (\rho)\, d\rho.
\eeq

     Evidently, $\cF_1(\infty) =0$ due to the asymptotic flatness conditions
     (\ref{flatas}).

     Consider $\cF_1 (h)$. The metric regularity at the horizon requires
     $r'(h) < \infty$, as is easily verified by calculating the Kretschmann
     scalar $R^{\alpha\beta\gamma\delta}R_{\alpha\beta\gamma\delta}$ (see,
     e.g., \cite{vac5}), therefore the quantity $\cF_1(h)$ is, in
     general, finite but nonzero. If, however, we admit a nonzero value of
     $A\phi'^2$ at $\rho=h$ whereas $A \sim \rho-h$, so that $\phi' \sim
     (\rho - h)^{1/2}$, then the integral in (\ref{iden}) will diverge at
     $\rho = h$ due to the second term in brackets in (\ref{F2}), and this
     will in turn lead to an infinite value of $\cF_1(h)$. We have to
     conclude that $A \phi'^2 \to 0$ as $\rho \to h$, and consequently
\[
     \cF_1(h) =  \frac{2r^2(h)}{r'(h)} V(h) \geq 0.
\]

     Thus the l.h.s. of \eq (\ref{iden}) is nonpositive and the r.h.s.
     is nonnegative. The only way to satisfy (\ref{iden}) is to put $V\equiv
     0$ and $\phi' \equiv 0$ in the whole range $\rho > h$, and the only
     solution for the metric is then \Sch.
$\DAL$

\medskip
     The above reasoning does not work if we admit $r'\leq 0$ at some
     $\rho\geq h$. Suppose that $r'=0$ at some $\rho_1 >h$ and $r'<0$
     at $h\leq \rho< \rho_1$. Then, even if $\phi'$ turns to zero together
     with $r'$, so that \eq (\ref{iden}) remains meaningful, the l.h.s. in
     (\ref{iden}), equal to $ -V(h)/r'(h)$, is positive in case $V(h) > 0$,
     and \eq (\ref{iden}) may hold. However, a horizon located at a \wh\
     throat, such that $r'> 0$ at $\rho > h$ and $r'(h) = 0$, is impossible.
     Indeed, we then obtain $-\cF_1 (h+0) = -\infty$ for $V(h) > 0$ and a
     nonpositive limiting value $-\cF_1 (h+0)$ in case $V(h)=0$.

\subsection{Numerical estimates}

     Using an observational restriction on the bulk energy scale $m_5$ and
     the well-known value of the 4D Planck energy, $m_4 = \kappa_4^{-1} =
     (8\pi G_{\rm N})^{-1/2} \approx 3.7\ten{18}$ GeV, we can estimate the
     density scale of brane matter (scalar field in our case) which can lead
     to a geometry drastically different from GR predictions.

     Namely, consider the conditions on a \wh\ throat: $r > 0$, $r'=0$ and
     $r'' \geq 0$. It is this phenomenon that restricts the validity of
     Theorem 3, to say nothing on the possible existence of \whs. As follows
     from \eq (\ref{01}), the quantity $A\phi'^2 - 2V > 0$ at the
     throat (we suppose $\phi'\ne 0$). Then, the l.h.s. of \eq (\ref{int})
     is simply $-r^{-2}$ at the throat, and the condition that the r.h.s. is
     negative yields (assuming, as before, $\Lambda_4=0$)
\beq                                                           \label{W}
      W := (A\phi'^2 - 2V)\Big|_{\rm throat}
      		\gsim W_- = \frac{24 \kappa_4^2}{\kappa_5^4}
			=  24 m_4^4 \biggl(\frac{m_5}{m_4}\biggr)^6,
\eeq
     where $m_4^4 \approx 0.75\ten{91}$ \dens\ is the Planck density. Note
     that $W/2 = -T_1^1$ is the radial pressure of the scalar field. The
     energy density is $T_0^0 = \half A\phi^2 + V$, and $T^0_0 > W/2$ if $V
     >0$.  Then $W/2$ can be used as a lower bound for $T^0_0$. If we even
     admit $V < 0$ such that $T^0_0 \ll W$ near the throat, the estimate of
     $W$ will still refer to radial pressure.

     A similar estimate can be obtained for a possible particlelike solution
     in case $V \geq 0$. Indeed, \eq (\ref{univ-b}) can only hold if the
     integrand is negative in some region. If there is no \wh\ throat (i.e.,
     the above estimate does not work) and $r'> 0$ in the whole space,
     this may happen at the expense of the last term $-2f^2$ which means
     that, at least, $f \equiv A\phi'^2 > 12 \kappa_4^2/\kappa_5^4 = \half
     W_-$. A particlelike solution with $V\geq 0$ thus requires a
     sufficiently high kinetic energy density $\half A\phi'^2 > \qua W_-$ in
     a certain region of space; due to $V \geq 0$, the full density $\half f
     +V$ will be still larger.

     The quantity (\ref{W}) is extremely sensitive to the value of $m_5$.
     Thus, it has been claimed on the basis of accelerator data that the
     bulk energy scale $m_5$ might be as small as a few TeV. Taking, for
     certainty, $m_5 = 10$ TeV, we obtain $m_5/m_4 \approx 2.7\ten{-15}$,
     and consequently $W_- \approx 7.3\ten 4$ \dens. This is smaller than
     the mean density of white dwarfs ($5 \div 6 \ten{5}$ \dens\
     \cite{allen}.

     A much tighter restriction follows from the recent short-range
     Newtonian gravity tests \cite{long}: the length scale
     $\ell = (6/|\Lambda_5|)^{1/2} \lsim 1$ mm.
     \eqs (\ref{La4}) with $\Lambda_4 =0$ then lead to
\beq                                                         \label{m5}
     \frac{m_5}{m_4} = (\pi\ell m_4)^{-1/3}
		   = \biggl(\frac{l_{\rm Planck}}{\ell}\biggr)^{1/3}
\eeq
     whence $(m_5/m_4)^3 \gsim 2.6\ten{-32}$, and $W \gsim 10^{29}$ \dens, an
     enormous value, many orders of magnitude over the nuclear density.

     We can conclude that as long as the scalar field density and pressure
     are much smaller than this value of $W$, \wh\ throats cannot appear,
     and all the above no-go theorems, known for a scalar field in GR,
     remain valid in a minimally coupled \bw.

\section {Concluding remarks}

     We have seen that some no-go theorems for scalar vacuum hold in GR but
     can be violated in a minimally coupled \bw. In particular, the no-hair
     theorem for \asflat\ \bhs\ holds only under the additional assumption
     that there is no \wh\ throat in the domain of outer communication.
     Accordingly, static scalar vacuum in a \bw\ can in principle yield as
     many as three kinds of solutions impossible in GR: (i) traversable
     \whs, (ii) particlelike solutions for scalar fields with $V\geq 0$ and
     (iii) black holes with scalar hair, whose horizons are located beyond
     \wh\ throats.  All of them become possible due to the strong field
     behaviour of brane-world gravity, more precisely, due to the properties
     of the tensor $\Pi\mN$. An estimate shows, however, that such exotic
     objects can only form by a scalar field whose energy density (and/or
     radial pressure) are many orders of magnitude larger than the density
     of an atomic nucleus ($\sim 10^{15}$ \dens).

     The above results can be readily generalized in two respects: for
     brane worlds of dimensions greater than four and for scalar field
     multiplets $\{\phi^a\}$ with Lagrangians of sigma-model type
\beq
     L_\sigma = h_{ab}(\phi)g\MN \d_\mu\phi^a \d_\nu\phi^b - V(\phi),
     	 		         			       \label{sigma}
\eeq
     where $h_{ab} (\phi)$ is a positive-definite matrix of functions of
     $\phi^a$.

     If we abandon the minimal coupling assumption $E\mN=0$, then a
     possible nonzero $E\mN$ brings about much ambiguity. Thus, in isotropic
     cosmology it adds ``dark radiation'' of arbitrary density (one extra
     constant) \cite{brane-rev,bra-scalar}. In static spherical symmetry it
     adds an arbitrary function of the radial coordinate
     \cite{br-kim,vis-wilt,bbh1}; in particular, vacuum \bhs\ and \whs\ may
     be obtained with {\it any\/} functions $g_{00}(r)$ satisfying the
     proper boundary conditions \cite{br-kim,bbh1}. This is the case in the
     simplest RS2-type \bw\ models \Ref{RS2}, possessing a single
     extra dimension, ${\mathbb Z}_2$ symmetry with respect to the brane and
     no matter in the bulk, to say nothing of more complex models. The
     latter may include scalar fields in the bulk \cite{sc-bulk}, multiple
     (at least two) branes \cite{2bra}, timelike extra dimensions, lacking
     ${\mathbb Z}_2$ symmetry, a 4D curvature term \cite{shta}, etc; further
     references may be found in the cited papers and the reviews
     \cite{brane-rev}. To improve the predictive power of \bw\ scenarios, it
     seems necessary to remove the ``redundant freedom'', applying
     reasonable physical requirements such as regularity and stability to
     complete multidimensional models.

     The recent paper by Anderson and Tavakol \cite{AT} (which appeared when
     this study was completed) shows that, even in the RS2 framework,
     different characterizations of the bulk lead to different forms of
     the effective 4D equations, though ``all characterizations are
     equivalent if one considers the entire brane-bulk system''. In our
     view, the 4D gravity equations on the brane, which determine the
     observable picture of the Universe, must also be unique in a given
     bulk-brane system, and their formulation (\ref{EE4}) seems to be preferable
     due to a clear distinction of the bulk Weyl tensor contribution $E\mN$
     and the tensor $\Pi\mN$ quadratic in $T\mN$, important in strong
     fields. Other formulations \cite{AT} probably mix these contributions.
     The problem evidently needs a further study.

\subsection*{Acknowledgements}

KB and SF acknowledge partial financial support from the RFBR Grant
01-0217312a, the Russian Ministry of Education and the Russian Ministry of
Industry, Science and Technologies.


\end{document}